\documentclass[aps,prapplied,reprint,groupedaddress,superscriptaddress,amsmath,amssymb]{revtex4-2}
\usepackage[T1]{fontenc}
\usepackage[utf8]{inputenc}
\usepackage{lmodern}
\usepackage[english]{babel}
\usepackage[colorlinks=true,linkcolor=blue]{hyperref}
\usepackage[capitalize,nameinlink]{cleveref}
\usepackage{siunitx}
\usepackage{graphicx}
\begin{document}

%\preprint{APS/123-QED}

\title{Tunable Josephson voltage source for quantum circuits}

\author{J.-L. Smirr}
\author{P. Manset}
\affiliation{JEIP, UAR 3573, CNRS, Collège de France, PSL University, 11, place Marcelin Berthelot, 75005 Paris, France}
\author{Ç. Ö. Girit}
\email{caglar.girit@cnrs.fr}
\affiliation{JEIP, UAR 3573, CNRS, Collège de France, PSL University, 11, place Marcelin Berthelot, 75005 Paris, France}
\affiliation{Quantronics Group, Université Paris Saclay, CEA, CNRS, SPEC, 91191 Gif-sur-Yvette, France}

\date{\today}

\begin{abstract}
Noisy voltage sources can be a limiting factor for fundamental physics experiments as well as for device applications in quantum information, mesoscopic circuits, magnetometry, and other fields.
The best commercial dc voltage sources can be programmed to approximately six digits and have intrinsic noise in the microvolt range.
On the other hand the noise level in metrological Josephson-junction based voltage standards is sub-femtovolt.
Although such voltage standards can be considered ``noiseless,'' they are generally not designed for continuous tuning of the output voltage nor for supplying current to a load at cryogenic temperatures.
We propose a Josephson effect based voltage \emph{source}, as opposed to a voltage standard, operating in the \qtyrange{30}{160}{\uV} range which can supply over \qty{100}{\nA} of current to loads at \unit{\milli\kelvin} temperatures.
We describe the operating principle, the sample design, and the calibration procedure to obtain continuous tunability.
We show current-voltage characteristics of the device, demonstrate how the voltage can be adjusted without dc control connections to room-temperature electronics, and showcase an experiment coupling the source to a mesoscopic load, a small Josephson junction.
Finally we characterize the performance of our source by measuring the voltage noise at the load, \qty{50}{\pV}~rms, which is attributed to parasitic resistances in the cabling.
This work establishes the use of the Josephson effect for voltage biasing extremely sensitive quantum devices.
\end{abstract}

\maketitle

\section{Introduction}

Low-noise voltage sources are essential for sensitive electronics and basic science.
The ultimate in voltage accuracy and precision is provided by quantum voltage standards based on the Josephson effect~\cite{jeanneret_application_2009}.
The parts-per-billion frequency stability of microwave oscillators combined with the ac Josephson relation, $V=(h/2e) \times f$, between voltage $V$ and frequency $f$, have enabled a metrological definition of the volt~\cite{mohr_codata_2002}.
Although Josephson standards are excellent for calibration or reference purposes~\cite{Rufenacht_Impact_2018}, certain limitations have prevented their use as tunable voltage sources for precision applications in fields such as quantum information and mesoscopic physics.
Most Josephson standards only provide a single fixed voltage in the range \qtyrange{1}{10}{\V}.
The programmable Josephson voltage standard allows adjustment of the reference voltage, but only in discrete steps with undefined transient voltages during changes~\cite{burroughs_nist_2011}.
The Josephson arbitrary waveform synthesizer allows fine voltage resolution but requires complicated pulse-generation circuitry~\cite{benz_pulsedriven_1996}.

Here, we demonstrate a tunable, low-noise, metrologically accurate voltage source~\cite{giritsmirr2023} that can operate continuously in the range \qtyrange{30}{160}{\uV} and which is simple to implement as it only requires a tunable microwave generator.
This voltage range is suitable for applications such as Josephson spectroscopy~\cite{griesmar_superconducting_2021}, topological transconductance quantization~\cite{peyruchat_transconductance_2021}, dc-pumped parametric amplification and squeezing~\cite{jebari_near-quantum-limited_2018,mendes_parametric_2019}, single microwave photon generation~\cite{grimm_bright_2019,rolland_antibunched_2019}, entangled beam generation~\cite{peugeot_generating_2021,ma_antibunched_2021}, and quantum thermodynamic engines~\cite{lorch_optimal_2018}.

\begin{figure}
  \includegraphics[width=\columnwidth]{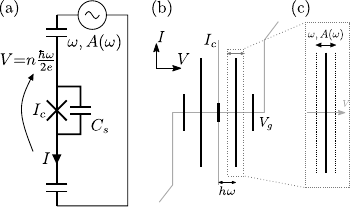}
  \caption{%Principle of Josephson tunable voltage source.
    (a) Circuit schematic of a Josephson tunnel junction driven by a microwave source showing locking to a Shapiro step at voltage $V = n\hbar\omega/2e$.
    Here, $I_c$ is the junction critical current and $C_s$ is a shunt capacitance.
    (b) Sketch of the current-voltage characteristic of the junction with (black) and without (light gray) microwave drive.
    When the drive amplitude is appropriately adjusted, Shapiro steps (thick vertical lines) appear.
    (c) After calibrating for the optimal microwave amplitude-frequency dependence $A(\omega)$ that maximizes the height of the first Shapiro step, the output voltage $V = \hbar\omega/2e$ can be continuously tuned without losing phase lock.
  }
  \label{fig:intro}
\end{figure}

The fundamental problem with using Shapiro steps for a tunable voltage source is that the steps are only stable for certain values of microwave power.
In a Shapiro voltage \emph{standard}, the amplitude of the microwave drive is precisely tuned at the working frequency to maximize the step height, thereby maintaining a stable output voltage.
Continuously changing the voltage implies changing the working frequency; however, without simultaneous adjustment of the microwave amplitude, the voltage can switch to an arbitrary Shapiro step.
By determining the optimal amplitude at each frequency, one can maintain a stable Shapiro step over a wide range of output voltage.
This is the basic principle of the Josephson tunable voltage source.

A schematic of the device is shown in~\cref{fig:intro}(a).
A Josephson tunnel junction of critical current $I_{c}$ is capacitively coupled to a microwave voltage source of frequency $\omega$ and amplitude $A(\omega)$.
The junction is shunted by a capacitance $C_{s}$, which lowers the plasma frequency, $\omega_{p} \approx \sqrt{2eI_c/\hbar C_s}$, allowing stable operation at lower voltages~\cite{kautz_noise_1996}.
When the drive amplitude $A(\omega)$ is appropriately adjusted for the working frequency $\omega$, the junction locks to the $n$th Shapiro step at voltage $V = n\hbar\omega/2e$.
% A direct connection is used to source the dc voltage to another device or for characterization.

The locking process can be understood from a sketch of the current-voltage ($IV$) characteristic of the Josephson tunnel junction, \cref{fig:intro}(b).
Without a microwave drive, the $IV$ characteristic (light gray) shows a supercurrent peak of amplitude $I_c$ at zero voltage; a ``subgap'' region of zero average current for $0 < |V| < V_g$; and the quasiparticle branch for $|V| > V_g$, the superconducting gap voltage, where one recovers the normal-state resistance~\cite{orlandoFoundationsAppliedSuperconductivity1991}.
Under microwave drive, the $IV$ characteristic shows Shapiro steps, vertical current peaks at voltages $V_n=n\hbar\omega/2e$ (thick black lines)~\cite{shapiro_effect_1964}.
When operated on a Shapiro step, the junction acts as a noiseless voltage source that can supply currents close to the maximum of the peak, $I_{n}(V_{n})$.

Although the step position $V_n$ depends solely on the microwave frequency and fundamental constants, the Shapiro step height depends on the microwave power at the junction, which varies with both the amplitude and frequency of the microwave source.
Attenuation and imperfections in the transmission lines connecting the source to the junction, as well as an impedance mismatch at the junction, will influence the power delivered to the junction.
Due to the Josephson nonlinearity~\cite{likharev_dynamics_1986}, the effective junction impedance depends strongly on frequency and drive amplitude for $\omega\lesssim\omega_{p}$.
At fixed frequency, the step height oscillates with microwave power, passing through zeros where phase locking of the voltage is not possible.

Because of the variability in coupled power with frequency, it is in general impossible to stay voltage locked while adjusting only the drive frequency.
If the drive frequency falls on a value where the step height is zero, the voltage will switch to another step at different $n$, to the subgap region, or to the quasiparticle branch.
To prevent unlocking, we propose a calibration procedure to determine the optimal power $A(\omega)$ to maximize the Shapiro step of a given order $n$.
As shown in \cref{fig:intro}(c), by adjusting the microwave power according to $A(\omega)$, the Shapiro step height is maintained, and therefore the output voltage of the source can be tuned in a stable and continuous manner.

\section{Implementation}

\begin{figure}
  \includegraphics[width=\columnwidth]{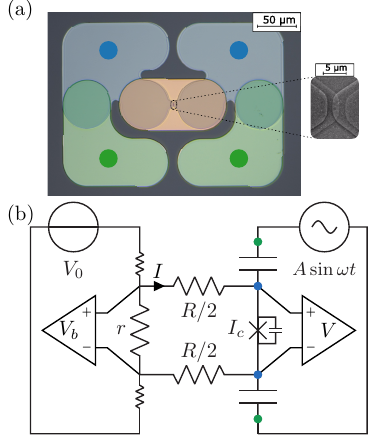}
  \caption{%Principle of Josephson tunable voltage source.
    (a) False-colored optical image of Josephson tunable voltage source and electron micrograph of the Josephson tunnel junction (inset).
    The shunt capacitor is highlighted in orange, and bonding pads are highlighted in blue (dc connections) and green (microwave bias connection).
    (b) Circuit schematic for measurement of device current-voltage characteristic.
    An applied dc voltage $V_0$ induces the current $I = (V_b-V)/R$ through the junction.
    Here, $V_b$ and $V$ are measured with differential voltage amplifiers and a microwave drive is applied via the signal generator $A\sin\omega t$.    
  }
  \label{fig:setup}
\end{figure}

An image of a microfabricated device is shown in \cref{fig:setup}(a), where the junction, capacitors, and pads have been colorized to match the circuit schematic in~\cref{fig:setup}(b).
The Josephson tunnel junction is made from aluminum using a Dolan-bridge technique and has a gap voltage of $V_{g} = 2\Delta/e = \SI{400}{\uV}$.
The shunt capacitor is a metal-insulator-metal structure with aluminum oxide dielectric and aluminum pads deposited by electron-beam evaporation.
The plasma frequency of the capacitively shunted junction is estimated to be $\omega_{p}=\qty{6.8}{\GHz}$ and the inferred shunt capacitance is $C_{s}\approx \qty{1.35}{\pF}$, which is consistent with the capacitor dimensions and dielectric constant.
Electrical connections for the dc source (blue) and ac drive (green) are made via wirebonds to the indicated pads.

The device is loaded into a custom broadband microwave sample holder designed for operation up to \SI{40}{\GHz} and cooled in a dilution refrigerator with a base temperature of approximately \SI{10}{\milli\K}.
All measurement lines are filtered to reduce electronic noise and the microwave lines are attenuated to limit radiation from room temperature.
The dc measurement and bias circuits are differential to reduce common-mode noise.
Details of the experimental setup can be found in~\cref{app:expt}.

Referring to the left-hand side of the schematic shown in \cref{fig:setup}(b), to measure the current-voltage ($IV$) characteristic, first a voltage $V_0$ is applied to the dc bias line.
The series combination of the resistors in the filtered bias line and a small shunt resistor $r \ll R$ located at the lowest-temperature stage of the cryostat forms a voltage divider.
The attenuated dc voltage $V_b \ll V_0$ across the shunt resistor is measured with a low-noise room-temperature differential amplifier.
A current $I$ flows through the Josephson junction and two series resistors $R/2$.
Another differential amplifier measures the dc voltage $V$ across the junction and the current $I$ is calculated as $(V_b-V)/R$.

The right-hand side of \cref{fig:setup}(b) depicts a microwave drive $A\sin\omega t$ produced by a room-temperature signal generator.
The microwaves are applied to the junction through microwave cabling and on-chip capacitors.
Measurement of the $IV$ characteristic of the junction consists of fixing $A$ and $\omega$, sweeping the dc bias $V_0$, and simultaneously measuring $V_b$ and $V$ to calculate $I$.
%As the junction response is nonlinear, bidirectional sweeps of $V_0$ in different ranges are necessary to reconstruct a full $IV$ characteristic.

\section{Calibration}

\begin{figure}
  \includegraphics[width=0.926\columnwidth]{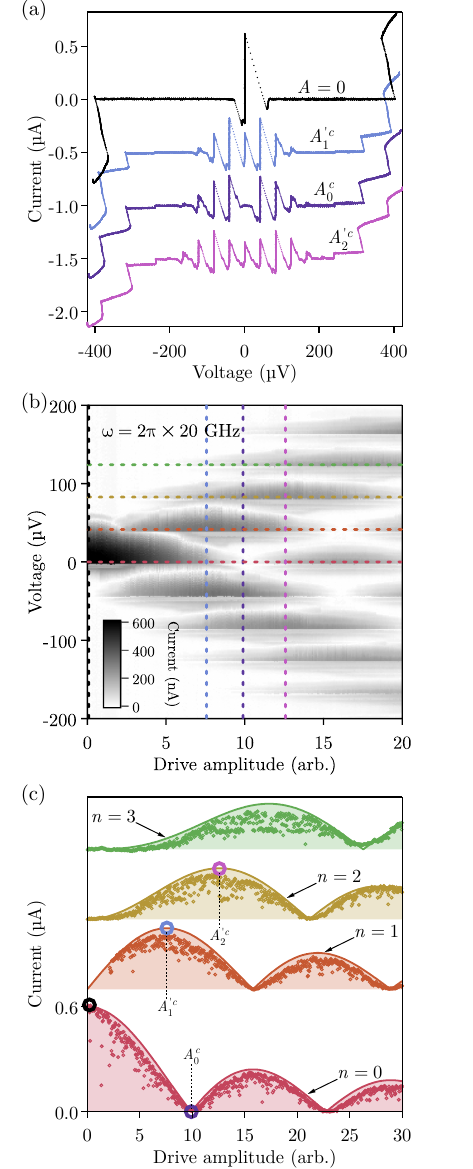}
  \caption{%Calibration of Josephson tunable voltage source.
    (a) Current-voltage characteristics of Josephson tunable voltage source  for different microwave drive amplitudes $A$ at fixed frequency $\omega_0/2\pi=\SI{20}{\GHz}$.
    Traces are offset by \SI{0.5}{\uA} for clarity.
    (b) Map of current as a function of microwave drive amplitude ($x$ axis) and voltage ($y$ axis) showing Shapiro step lobes.
    (c) Amplitudes of the supercurrent peak and first three Shapiro current steps, as extracted from the current-voltage map in (b), are plotted as a function of drive amplitude.
}
  \label{fig:data}
\end{figure}

\cref{fig:data}(a) shows the $IV$ characteristics as a function of microwave power at a fixed drive frequency $\omega_{0}/2\pi=\SI{20}{\GHz}$.
In the absence of microwave drive, $A=0$, there is a single peak at zero voltage, the supercurrent peak (black trace).
The mean switching current is \SI{600}{\nano\ampere}, close to the value of the current at the quasiparticle knee near the gap voltage $V_g = \qty{400}{\uV}$, as expected by the Ambegaokar-Baratoff theory~\cite{griesmar_superconducting_2021}.
There are no Shapiro steps, and the subgap region is flat up to the gap voltage $V_g$.
The bias voltage $V_0$ is swept upwards, and ``switching'' between the supercurrent peak and the subgap region is deduced from the dotted diagonal line near zero voltage.
The slope of this load line is given by $1/(R+r)$.

%.
% This value is deduced from comparing the geometry, fabrication parameters and critical currents to reference samples where $\omega_p$ was determined from microwave response.
%  the frequency at which the junction $IV$ characteristic is most sensitive to microwave drive
% At resonant drive, $\omega\approx\omega_{p}$, the $IV$ is strongly distorted even at extremely small power levels.
% (area \qty{3050}{\micro\metre\squared}, thickness \qty{100}{\nano\metre}).

Shapiro steps---current peaks in the subgap region---appear for nonzero drive amplitude $A > 0$ (colored traces).
At an appropriate microwave drive strength, corresponding to a critical amplitude $A_{1}^{'c}$, the height of the first-order peak at $V=\hbar\omega_{0}/2e=\SI{41.4}{\uV}$ is maximal and approximately~\SI{250}{\nA}.
Increasing the microwave amplitude to $A(\omega_{0}) = A_{0}^{c}$, the supercurrent peak ($n=0$) vanishes but steps of higher order persist.
With stronger microwave drive, steps of higher order can be maximized, such as the second one ($A_{2}^{'c}$).
In this notation, at drive amplitude $A_{n}^{'c}$, the $n$th current peak is at a maximum, whereas at $A_{n}^{c}$ the $n$th peak is at its first nontrivial zero.

In~\cref{fig:data}(b), the $IV$ characteristics are plotted as a two-dimensional map over a broad range of microwave powers, with each pixel corresponding to the maximal absolute value of the measured current at a given bias voltage and drive amplitude.
Shapiro steps appear as dark gray points grouped in horizontal line segments, with the contrast indicating the step height.
The periodic, triangular pattern is expected from the Bessel-function dependence of step heights~\cite{kautz_noise_1996}.
The $IV$ traces of~\cref{fig:data}(a), measured at constant power, are indicated by vertical dashed lines.

The heights of steps $n=\numrange{0}{3}$, determined by taking the corresponding cuts of~\cref{fig:data}(b) (horizontal dashed arrows), are plotted as a function of microwave power in~\cref{fig:data}(c).
They are fit to Bessel functions of the first kind, $|J_n|$, for orders $n=\numrange{0}{3}$.
In contrast to~\cref{fig:data}(a), here, we see the continuous evolution of the step heights with $A(\omega)$: the supercurrent peak decreases and reaches its first zero ($A_{0}^{c}$) while the first- and second-order peaks reach their respective maxima at $A_{1}^{'c}$ and $A_{2}^{'c}$.

In principle, any of the critical points $A_{0}^{c}(\omega)$,$A_{1}^{'c}(\omega)$, or $A_{2}^{'c}(\omega)$ can be used to calibrate the incident power at the junction for frequencies $\omega\gtrsim\omega_{p}$, where the Bessel-function dependence is valid~\cite{kautz_noise_1996}.
Junctions biased at lobe maxima, such as $A_{1}^{'c}(\omega)$ and $A_{2}^{'c}(\omega)$, serve as square-law power detectors whereas junctions biased at zeros such as $A_{0}^{c}(\omega)$ serve as linear detectors~\cite{richards_millimeter_1973, zmuidzinas_superconducting_2004}.
In addition to calibrating the microwave power using the Josephson effect, photon-assisted tunneling (PAT) of quasiparticles near the gap voltage can also be used for calibration~\cite{tucker_quantum_1985}.

For simplicity, we either use the first zero of the supercurrent, $A_{0}^{c}(\omega)$, or the PAT current to calibrate the power for a frequency range spanning approximately \qtyrange{8}{40}{\GHz}.
Both of these techniques avoid complications arising from the hysteretic, nonlinear shape of the junction's $IV$ characteristic.
At each frequency, we determine the power necessary to either zero the supercurrent peak or to obtain a reference value of the PAT current near the quasiparticle knee.
From the Bessel-function dependence of the critical point or the PAT current, we calculate the drive power necessary to maximize the Shapiro step of a given order.
Technical details of the calibration procedure are given in~\cref{app:calib}.

\section{Performance}

\begin{figure}
\includegraphics[width=0.968\columnwidth]{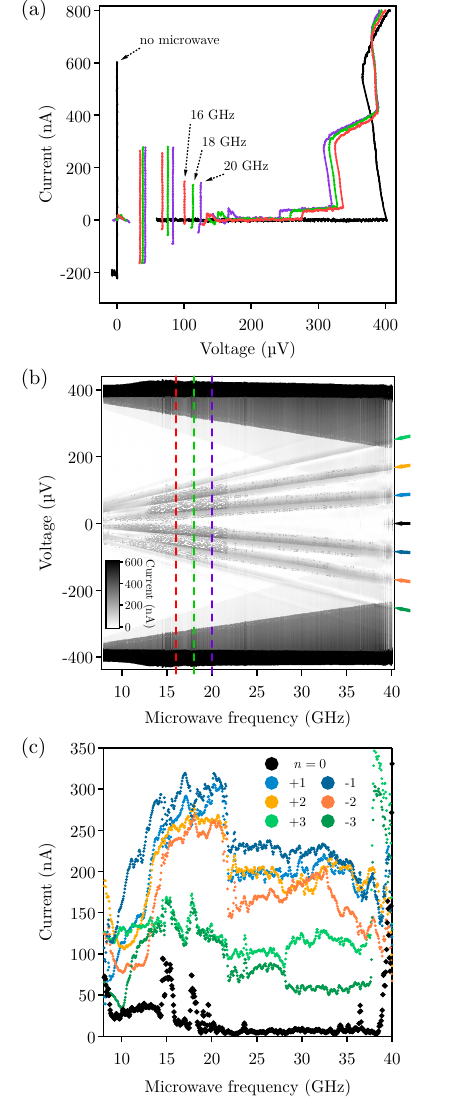}
    \caption{
      (a) Current-voltage characteristics measured at optimal microwave amplitude $A_{0}^{c}(\omega)$ minimizing the supercurrent for $\omega/2\pi=\qtylist[list-units=single]{16;18;20}{\GHz}$.
      A reference $IV$ without microwave drive is shown in black.
      Data is measured with a positive sweep in bias voltage, resulting in smaller negative peaks.
      (b) Current voltage map measured at $A_{0}^{c}(\omega)$ plotted as a function of microwave drive frequency.
      The $IV$ characteristics in (a) are indicated by vertical dashed lines.
      (c) Height of Shapiro peaks of order $n$ extracted from (b) along the lines indicated by arrows.
    }
  \label{fig:operation}
\end{figure}

To demonstrate the quality of the calibration, \cref{fig:operation}(a) plots current-voltage characteristics with a microwave drive at \qtylist[list-units = single]{16;18;20}{\GHz} and amplitudes $A_{0}^{c}(\omega)$ corresponding to the first zero of the supercurrent.
Shapiro steps up to order $n=4$ are visible at multiples of voltages \qty{33.09}{\uV} for \qty{16}{\GHz} drive, \qty{37.22}{\uV} (\qty{18}{\GHz}), and \qty{41.36}{\uV} (\qty{20}{\GHz}), with a corresponding step spacing $n \times \qty{2}{\GHz} \times h/2e \approx n \times \qty{4.14}{\uV}$.
As anticipated in the schematic,~\cref{fig:intro}(c), the Shapiro step height within each order is approximately the same.
A figure of merit $\mu$ for calibration is the ratio of the residual mean switching current, which should be zero for perfect calibration, to the maximum switching current, which is approximately \qty{600}{\nA} in the reference $IV$ without microwave drive (black trace).
The residual mean switching current for all three frequencies in~\cref{fig:operation}(a) is less than~\qty{20}{\nA}, giving $\mu \approx 3\%$.

\cref{fig:operation}(b) shows a current-voltage map when driving at amplitude $A_{0}^{c}(\omega)$ over a wide frequency range.
The measured average voltage is shown on the left-hand axis, applied microwave drive frequency is on the bottom axis, and the maximum absolute value of the measured current is represented by the pixel contrast.
The $IV$ plots of \cref{fig:operation}(a) at \qtylist[list-units = single]{16;18;20}{\GHz} are indicated by color-coded dashed vertical lines.
Colored arrows along the right-hand axis indicate Shapiro steps of order $n=-3$ to $n=3$.
Since the microwave drive is adjusted to $A_{0}^{c}(\omega)$, the supercurrent peak along the horizontal axis is virtually absent, and the nonzero orders are easily visible, although none is maximized.

From the current-voltage map, we numerically extract the mean switching current of the Shapiro steps and plot them as a function of frequency [\cref{fig:operation}(c)].
The residual supercurrent (black trace) is low, with $\mu \lesssim 10\%$ for almost the entire frequency range.
In addition, the height of the first-order steps ($n=\pm1$, blue traces) is more than~\qty{200}{\nA}, or approximately a third of the zero-drive supercurrent, for a large range of frequencies.
The Bessel-function dependence of step heights predicts that the first-order steps should be approximately $0.519$ times the supercurrent peak for a microwave drive $A_{0}^{c}(\omega)$, and this appears to be the case in the frequency range \qtyrange{15}{20}{\GHz}, where the peak height reaches \qty{300}{\nA}.
The measured step heights for the higher orders also agree with theory, as the expected peak-height ratios $I_{n}/I_{c}$ for $n = 2 \text{ and } 3$ are $0.432$ and $0.2$, respectively, compared to $1/3$ ($n=\pm2$, orange traces) and $1/6$ ($n=\pm3$, green traces) estimated from the data.

Some of the measured variation in the peak heights in~\cref{fig:operation}(c) can be explained, such as the sharp drop above approximately~\qty{22}{\GHz}, which is attributed to the deactivation of an attenuator, resulting in increased spurious noise, lowering switching currents.
Discrepancies at lower frequencies may be due to weaker applicability of the Bessel-function form for step heights in that frequency range.
Furthermore, microwave resonances on the chip or in the sample box may give rise to the narrow features superimposed on a generally smooth step-height dependence.

To use the device as a tunable voltage source it is crucial that the step amplitudes do not dip toward zero anywhere in the frequency range.
Such dips will reduce the maximum possible source current and in the worst case, destroy phase lock.
A load that sinks a current greater than the step amplitude will cause the source to switch off the Shapiro step.
With our calibration procedure, it is possible to source at least \qty{50}{\nA} using any order over almost the entire frequency range, with the only exception being a narrow region near \qty{10}{\GHz} at $n=-3$.

\section{Stability}

Next, we consider the possibility of voltage locking without dc bias.
Shapiro steps that cross at zero current---as shown in the sketch in \cref{fig:intro}(b) and demonstrated in the $IV$ characteristic in \cref{fig:operation}(a)---may occur in underdamped Josephson junctions~\cite{levinsen_inverse_1977} and are used in zero-bias voltage standards~\cite{kautz_proposed_1980,kautz_noise_1996}.
A nonzero voltage-locked state is stable even in the absence of a bias current, and a voltage $n\hbar\omega/2e$ for $|n|>1$ can spontaneously develop across a microwave-driven junction.

\begin{figure*}
\includegraphics[width=\textwidth]{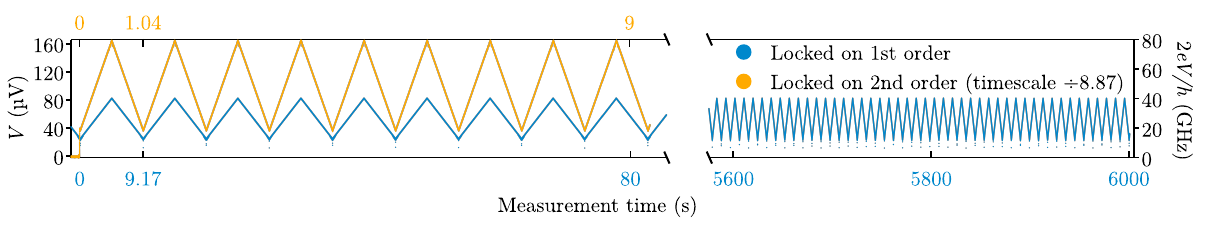}
\caption{
      After switching to the first Shapiro step (blue), the frequency is modulated in the range \qtyrange{10}{40}{\GHz} with a triangular waveform having a period of \qty{9.17}{\s} (bottom time scale).
      The power is adjusted according to the calibration to maintain the step amplitude.
      The measured junction voltage is continuous and follows the Josephson relation $V=\hbar\omega/2e$ without switching to other steps over a total time of \qty{100}{\min}.
      In orange the junction is locked to the second-order step at $V=2\hbar\omega/2e$, the sweep period is \qty{1.036}{\s}, and the measurement duration is approximately \qty{9}{\sec} (top time scale).
      The biasing circuit is disconnected with a cryogenic switch during the measurement (see main text).
    }
  \label{fig:locking}
\end{figure*}

In~\cref{fig:locking}, we show how the dc voltage on our junction can spontaneously lock to the drive frequency without any dc bias and then remain locked as the voltage is tuned continuously over a large range.
For this purpose, we use a cryogenic switch to disconnect the dc bias $V_0$, shunt resistor $r$, and voltage probe $V_b$ on the left-hand side of \cref{fig:setup}, breaking the current loop and forcing $I=0$ (more details are given in \cref{app:expt} and \cref{fig:figSup-SourceEmitter}); however the junction voltage $V$ can still be measured with dc connections to the voltage amplifier.
To spontaneously switch to a Shapiro step, the phase must first unlock from the supercurrent branch at $n=0$.
To do so, the amplitude is ramped to the calibration point $A_{0}^{c}(\omega)$ where the effective supercurrent is zero.
At this point, as a result of fluctuations, the voltage jumps to a Shapiro step of order $n$, which is typically $\pm 1$.
We then set the power to the calibration value $A_{n}^{'c}(\omega)$, which maximizes phase stability, subsequently adjusting it in tandem with the frequency.

We ramp the frequency up and down while locked to the first, $n=1$, or second Shapiro step, $n =2$, for a total duration of \qty{100}{\min} ($n=1$) or \qty{9}{\second} ($n=2$). 
The curve is continuous, and the voltage does not switch to zero or to another step.

\section{Noise}

\begin{figure}
  \includegraphics[width=\columnwidth]{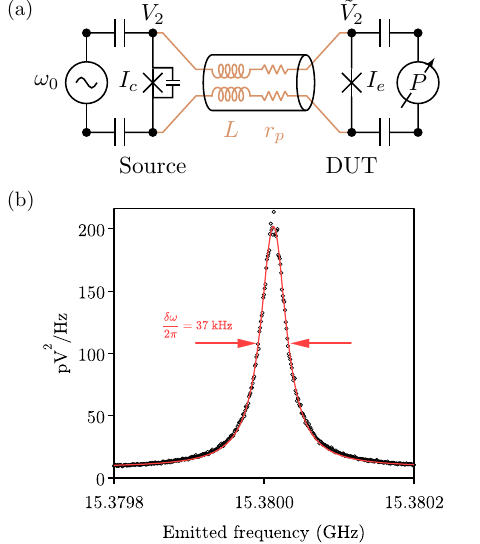}
  \caption{
    (a) A small Josephson junction (critical current $I_e < I_c$), serving as a device under test (DUT, right), is coupled to the Josephson voltage source via a superconducting twisted-pair cable of inductance $L$ and small parasitic resistance $r_p$.
    The source and DUT are located in separate sample boxes~(\cref{fig:figSup-SourceEmitter}).
    A microwave drive of frequency $\omega_{0}/2\pi=\qty{7.690}{\GHz}$ is applied to switch the junction to the second Shapiro step at voltage $V_{2}=2\hbar\omega_{0}/2e=\qty{31.80}{\uV}$.
    The dc voltage at the DUT, $\tilde{V}_2\approx V_2$, is converted by the Josephson effect to a microwave signal of frequency $\omega_{m}\approx2\omega_{0}$ which is measured with a spectrum analyzer (symbol $P$) after amplification (not shown, see \cref{fig:figSup-CryostatWiring}).
    (b) The power spectrum shows a narrow peak at $\omega_{m}/2\pi=\qty{15.38001260(2)}{\GHz}$ with a FWHM linewidth of~\qty{37}{\kHz} (red fit).
  }
  \label{fig:noise}
\end{figure}

To use the Josephson source in an experiment requiring a noiseless, tunable dc voltage, the device under test (DUT) should be connected to the source via a low-resistance, high-inductance interconnect, as shown in the circuit~\cref{fig:noise}(a).
Any parasitic resistance $r_p$ between source and DUT should be as small as possible to limit thermal noise.
The inductance $L$ serves as an rf choke, blocking the microwave drive at $\omega_{0}$ as well as harmonics $n\omega_{0}$ generated by the source.
Ideally, the interconnection between the source and the DUT should be designed to avoid resonances at frequencies in the desired voltage range.
The source and DUT can be located in the same enclosure, taking care to avoid microwave leakage between the source and DUT compartments, or in separate sample boxes.

As a proof-of-concept demonstration, we connect the source to a DUT, which is itself a small Josephson tunnel junction of critical current $I_e \approx \qty{140}{\nA} \approx I_{c}/6$ [\cref{fig:noise}(a)].
The source and emitter are located in separate enclosures and connected by a superconducting twisted-pair cable (\cref{fig:figSup-SourceEmitter}).
Due to the ac Josephson effect this ``emitter'' DUT converts the dc voltage from the source to a microwave signal.
We couple the emitter via its on-chip capacitors to a cryogenic amplifier (see \cref{fig:figSup-CryostatWiring}) and measure this signal using a microwave spectrum analyzer.
Since the measured signal linewidth is proportional to the dc voltage noise at the emitter, we can characterize the residual voltage noise of our setup.
The Josephson emission linewidth has been directly measured in this fashion for both large~\cite{yanson_experimental_1965} and small junctions~\cite{hofheinz_bright_2011,cassidy_demonstration_2017} biased with a conventional, noisy, dc voltage source.
Effectively, the source junction down-converts a microwave drive to a dc voltage, which is then transmitted via low-frequency wiring to the emitter junction, which then up-converts the dc back to a microwave signal.

The microwave drive frequency is $\omega_{0}/2\pi=\qty{7.69}{\GHz}$ and the amplitude has been chosen to maximize the step of order $n=2$.
The source junction is dc biased on the second Shapiro step at $V_{2}=2\hbar\omega_{0}/2e$.
This allows differentiating the emitter Josephson frequency $2\omega_{0}$ from the drive frequency $\omega_{0}$ of the source junction.
Although the second harmonic of the microwave drive generated by the Josephson nonlinearity of the source junction is also at $2\omega_{0}=\qty{15.38}{\GHz}$, it is heavily attenuated by the cable connecting the source to emitter.
This ensures that we measure the microwave signal resulting from down-conversion to dc and subsequent up-conversion by the emitter instead of direct frequency doubling by the source.
We confirm that there is no harmonic leakage by checking that the microwave power output of the emitter at $2\omega_0$ is independent of the power incident on the source junction, provided that the source junction is locked on the second Shapiro step.
A direct frequency-doubling process by the source junction would yield a Bessel-function dependence of the power detected at $2\omega_0$, whereas up-conversion from dc by the emitter yields a microwave current of amplitude bounded by $I_e$, independent of the incident power.

\cref{fig:noise}(b) shows the spectrum measured at the cryostat base temperature.
The full-width at half maximum (FWHM) of the Lorentzian peak is \qty{37}{\kHz}.
The linewidth broadening, relative to that of the microwave source, can be explained by a small resistance $r_p$ between the source and the emitter, indicated in the schematic of \cref{fig:noise}(a).
The peak position $\omega_{m}/2\pi=\qty{15.38001260(2)}{\GHz}$ is detuned from $2\omega_{0}$ by a $\Delta\omega = 2\omega_{0}-\omega_{m}\approx2\pi\times\qty{12.6}{\kHz}$.
% The fit can be found in data/EmissionSpectrum.pxp chi-square of 3.00e-11
This detuning can result from a parasitic dc current $i_p$ flowing through the resistance $r_p$, which gives a voltage drop $i_p r_p= \hbar\Delta\omega/2e = V_2-\tilde{V}_2$.
The origin of this dc current is likely to be inelastic Cooper-pair tunneling~\cite{ingold_1992,holst_1994}, in which the microwave power generated by the emitter and absorbed in its environment is balanced by the dc power supplied by the source.

We obtain $r_p\approx\qtyrange{20}{30}{\milli\ohm}$ using three different methods: equating $r_p$ with the known contact resistance of the connectors (\qty{30}{\milli\ohm}); using $r_p= \hbar\Delta\omega/2ei_p$ with $i_p$ estimated as the inelastic Cooper-pair current arising from a \qty{50}{\ohm} microwave environment (\qty{20}{\milli\ohm}); and extracting $r_p$ from the slope of the detuned voltage $\tilde{V}_2$ as a function of a current bias applied to the source (\qty{24}{\milli\ohm}).

With zero series resistance between the source and the emitter, we would expect a much narrower linewidth than the measured value, \qty{37}{\kHz}.
The emission linewidth is broadened by thermal fluctuations in the resistor $r_p$, which have a voltage-noise density $4 k_B T r_p$ where $k_B$ is Boltzmann's constant and $T$ is the electronic temperature.
For a Lorentzian lineshape, the expected FWHM is $4\pi k_B T r_p (2e/h)^2$, assuming a rectangular power spectrum for the noise~\cite{stewart_power_1954,hofheinz_bright_2011}.
This result is independent of the cutoff frequency, which could be thermal, $k_B T/h$, or given by $1/r_p C$ for sufficiently large $r_p C$, where $C$ is the effective capacitance of the cables and wiring in parallel with the emitter junction.

%%%%% Calculation of linewidth
%
% ddg 4*pi*boltzmann constant*(50 millikelvin)*(25 milliohm)/(magnetic flux quantum)^2 in kilohertz !g
% https://www.google.com/search?hl=en&q=4*pi*boltzmann%20constant*(50%20millikelvin)*(25%20milliohm)%2F(magnetic%20flux%20quantum)%5E2%20in%20kilohertz
%
%%%%% Calculation of electronic temperature
%
% ddg (37 kHz)/(4*pi*boltzmann constant*(25 milliohm)/(magnetic flux quantum)^2) in millikelvin
%
% https://www.google.com/search?hl=en&q=(37%20kHz)%2F(4*pi*boltzmann%20constant*(25%20milliohm)%2F(magnetic%20flux%20quantum)%5E2)%20in%20millikelvin
%
%%%%% Calculation of broadening per milliohm of parasitic resistance at 25 mK
%
% ddg 4*pi*boltzmann constant*(25 millikelvin)*(1 milliohm)/(magnetic flux quantum)^2 !g
%
% https://www.google.com/search?hl=en&q=4*pi*boltzmann%20constant*(25%20millikelvin)*(1%20milliohm)%2F(magnetic%20flux%20quantum)%5E2%20in%20kilohertz

Using $r_p = \qty{25}{\milli\ohm}$, the emission linewidth is compatible with a reasonable electronic temperature, $T = \qty{36}{\milli\kelvin}$.
Reducing the linewidth to sub-\unit{\kHz} levels would require care to eliminate parasitic resistance between the source and the DUT.
Every milliohm of resistance at a low electronic temperature of \qty{25}{\milli\kelvin} contributes \qty{1}{\kHz} to the linewidth.
When the source and the DUT are separated by superconducting wires, the resistance of connector pins, which dominates in this case, as well as solder and printed-circuit-trace resistances must be minimized.
Despite the inadvertent parasitic resistance, the measured frequency precision is $\qty{2.4}{ppm}$.
The inferred low-frequency voltage-noise density at the emitter junction is approximately \qty{220}{\femto\volt\per\sqrt\Hz}, and the total rms voltage noise is less than \qty{50}{\pV}.

%%%%% Calculation of voltage noise density
%
% ddg sqrt((1 Hz)*(37 kHz)/pi*(magnetic flux quantum)^2) in femtovolts !g
%
% https://www.google.com/search?hl=en&q=sqrt((1%20Hz)*(37%20kHz)%2Fpi*(magnetic%20flux%20quantum)%5E2)%20in%20femtovolts 
%
%
%%%%% Calculation of total voltage noise
%
% assumes bandwidth is less than FWHM
%
% ddg sqrt((37 kHz)*(37 kHz)/pi*(magnetic flux quantum)^2) in picovolts !g
%
% https://www.google.com/search?hl=en&q=sqrt((37%20kHz)*(37%20kHz)%2Fpi*(magnetic%20flux%20quantum)%5E2)%20in%20picovolts

\section{Conclusion}

We have demonstrated a tunable, low-noise voltage source based on the ac Josephson effect, operating in the range \qtyrange{30}{160}{\uV} and capable of delivering \qty{100}{\nA} of current, making it well suited to loads in the \unit{\kilo\ohm} range.
The calibration procedure optimizes stability of the voltage output across a wide frequency range, and the source maintains phase lock without an applied dc bias for an indefinite period.
One can connect to the voltage source with low-frequency superconducting wires or interconnects, taking care to minimize parasitic resistances, but it is not necessary to have the source and the load in the same enclosure.
In practice, the dc connection used to monitor the output voltage can be eliminated given the metrological relationship of source voltage to microwave drive frequency.
% Without the dc connection, any additional low-frequency phase noise , and avoiding such a  specific to dc instrumentation like 1/f noise.
In addition to continuous tunability, the source has low noise and metrological accuracy.

Future work could extend the upper voltage limit, possibly using series arrays of Josephson junctions or employing different junction materials such as niobium.
To achieve lower operating voltages one would need to increase the shunt capacitance, which reduces the plasma frequency and lowers the minimum voltage for stable Shapiro steps~\cite{kautz_noise_1996}.
A more complex microwave drive, such as biharmonic or pulsed, could allow deterministic switching to specific Shapiro steps.
The current range could be extended using larger junctions or parallel arrays.
%Improving the microwave coupling could also help reaching stability at voltages closer to zero.

Our proof-of-concept experiment coupling the source to a small Josephson junction demonstrates the possibility of integrating ultralow-noise, tunable, cryogenic dc voltage bias into devices for quantum information, quantum sensing, and mesoscopic physics.
Specific applications include microwave photon emitters~\cite{hofheinz_bright_2011,peugeot_generating_2021}, Josephson spectroscopy~\cite{griesmar_superconducting_2021, peugeotTwotoneSpectroscopyHighfrequency2024d}, dc-pumped parametric amplification~\cite{mendes_parametric_2019,albertMicrowavePhotonNumberAmplification2024,martel}, and qubit stabilization~\cite{danner}.

\begin{acknowledgments}
  We thank V. Benzoni, J. Griesmar, L. Peyruchat, J.-D. Pillet, and F. Lafont for discussions and assistance with the experiments.
  We thank Fabien Portier and Ambroise Peugeot for critical feedback.
  This project has received funding from the European Research Council (ERC) under the European Union's Horizon 2020 research and innovation program (Grant Agreement No. 636744).
  The research was also supported by IDEX Grant No. ANR-10-IDEX-0001-02 PSL, a Paris ``Programme Emergence(s)'' Grant, and the CNRS ``Prématuration'' program.
\end{acknowledgments}

\section*{Data availability}

The data that support the findings of this article are openly available~\cite{smirr_2024_14577236}.

\appendix

\section{\label{app:expt} Experimental Details}

\begin{figure}
  \includegraphics[width=0.94\columnwidth]{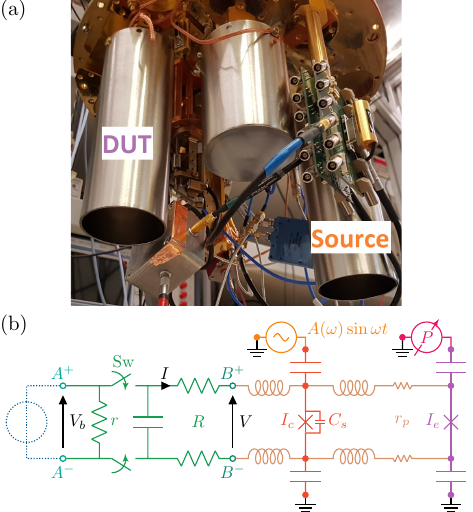}
  \caption{(a)~Photograph of the source (right) and DUT (left) in their magnetically shielded sample holders attached to the dilution refrigerator.
    The source, the DUT, and a biasing circuit box (not visible) are connected to each other by NbTi twisted pairs terminated by LEMO.00.302 connectors (not visible).
    Some elements appear on the photograph that are not related to the experiment described in this article.
    (b)~Schematic of the source (orange) connected to the emitter (purple) with biasing circuit (green) via NbTi twisted pairs (brown). Details of ac bias (gold) and measured ac power (pink) are available in~\cref{fig:figSup-CryostatWiring}, as well as details of dc supply (blue) and differential measurement of $V_b$ and $V$ (probe points indicated by circles $A^\pm$ and $B^\pm$). The DPDT electromechanical relay (Sw) allows the supply to be disconnected and the current shunt to be broken.}
  \label{fig:figSup-SourceEmitter}
\end{figure}

The experiments were conducted in a Bluefors LD250 dilution refrigerator with a base temperature below \qty{10}{\milli\K}.
A photograph and a complete simplified circuit can be found in~\cref{fig:figSup-SourceEmitter}.
%The voltage $V$, assuming no parasitic resistance, is equal to the voltage accross the source junction (and the emitter junction).

\begin{figure}
  \includegraphics[width=0.99\columnwidth]{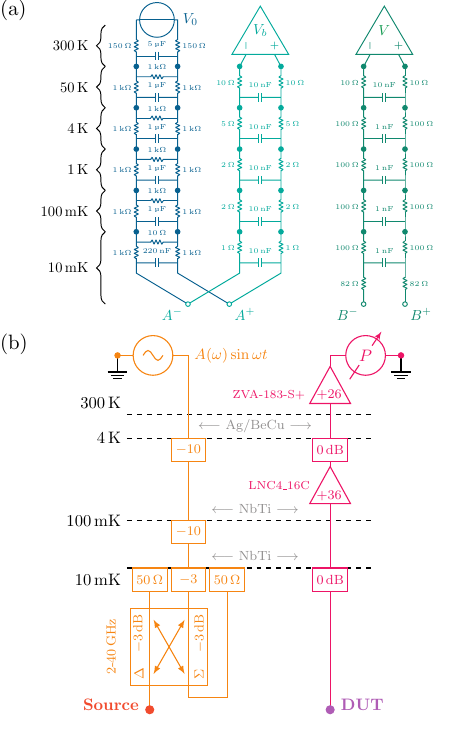}
  \caption{Details of the cryostat wiring:
    (a) the dc wiring comprises twisted pairs equipped with filtering $RC$ circuits thermally anchored at all temperature stages of the cryostat; 
    (b) the ac wiring uses standard copper coaxial cables with 2.92mm connectors, except NbTi between $\qty{10}{\milli\kelvin}$ and $\qty{4}{\kelvin}$ and silver-plated BeCu from $\qty{4}{\kelvin}$ to room temperature.
    The $\qty{1}{\kelvin}$ and $\qty{50}{\kelvin}$ stages are bypassed in ac.}
  \label{fig:figSup-CryostatWiring}
\end{figure}

The source $IV$ characteristics are obtained by sweeping $V_b$ with the double-pole, double-throw (DPDT) electromechanical switch Sw closed (KEMET EC2-12SNU)~\cite{beev_note_2012}.
The voltage $V$ is measured, and the current $I$ is calculated using $I=(V-V_b)/R$, where $R=\qty{80}{\ohm}$ is the sum of two series thin-film resistors.
Thermal and instrument noise is filtered as shown in~\cref{fig:figSup-CryostatWiring}, where the colors match those in \cref{fig:figSup-SourceEmitter}(b).

All connections are made with balanced, differential twisted pairs.
The \qty{20}{\ohm} bias resistor $r$ has a large volume in order to thermalize electrons.
A room-temperature floating voltage source supplies a voltage $V_0$, which is attenuated ($\sim 1/25000$) and $RC$ filtered to produce the bias voltage $V_b$.
Both $V_b$ and $V$ are measured using low-noise differential preamplifiers (NF LI-75).

During zero-bias operation, the supply can be disconnected by toggling Sw using a $\pm\qty{0.3}{\volt}$, \qty{0.1}{\second} long pulse delivered via a superconducting twisted pair (not depicted). A bypass capacitor, together with resistors $R$, filters unwanted transients.

The bias, source and emitter are interconnected by NbTi twisted pairs about \qty{30}{\cm} long, providing inductance to attenuate high frequencies while having zero dc resistance.
The residual parasitic resistance $r_p$ discussed in the main text is attributed to the pins of the connectors in the cable between the source and the emitter (LEMO 0.5 mm pins).

On the ac end of the source, the microwave drive is applied via coaxial cables with distributed attenuators totaling \qty{23}{\dB}.
The \qtyrange{2}{40}{\GHz} hybrid coupler adds \qty{3}{\dB} attenuation but terminates the sample with a \qty{50}{\ohm} load at \qty{10}{\milli\kelvin} while dumping low-frequency noise to another \qty{50}{\ohm} load.

On the ac end of the emitter, the spectrum is measured with a signal analyzer (Rohde\&Schwarz FSVA30) after cryogenic (LNF LNC4\_16C) and room-temperature (Mini-Circuits ZVA-183-S+) amplification.
Two ``\qty{0}{\dB}'' XMA attenuators help with cable thermalization.
No other filtering is used.

\begin{figure}
  % \vspace{2em}
  \includegraphics[width=0.94\columnwidth]{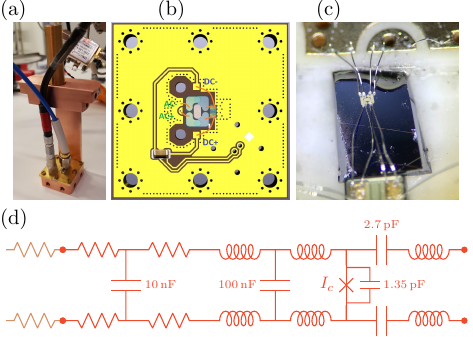}
  \caption{(a) Photograph of the sample holder used for both the source and the emitter.
    Sample holders are housed in cylindrical shields visible in~\cref{fig:figSup-SourceEmitter}(a)
    (b) Drawing of the PCB and wirebonds to the source sample, with \qty{100}{\nF} silicon and \qty{10}{\nF} ceramic filtering capacitors.
    (c) Close-up photo of the sample wirebonded to the \qty{2.92}{\mm} PCB launch and the \qty{100}{\nF} silicon filtering capacitor.
    Wirebonds to dc pads are partially visible.
    (d) Electrical schematic of the source chip and PCB. The emitter schematic is identical, but the $\qty{1.35}{\pF}$ on-chip shunt capacitor $C_s$ is absent.
    The inductors represent Al wirebonds, and the resistors represent the nonsuperconducting elements, including cable connector (brown), socket, and PCB tracks, which comprise the parasitic resistance $r_p$.
  }
  \label{fig:figSup-ParasiticResistance}
\end{figure}

The sample is mounted on a printed circuit board (PCB) covered by a copper lid used as a mechanical and thermal anchor to the mixing chamber of the cryostat [ \cref{fig:figSup-ParasiticResistance}].
To minimize the effect of electromagnetic cavity resonances, the sample box is designed to be small, and an Eccosorb absorber is affixed to the lid. Filtering capacitors are added to damp electrical modes in wirebonds and the PCB.
The sample is magnetically shielded by the combination of an inner aluminum cylinder and an outer cylinder made of Cryophy, a high-magnetic-permeability material that is compatible with low temperatures.
The PCB includes sockets for a differential dc connector (LEMO~00.302) and a microwave connector (2.92mm). 

The Dolan-type aluminum Josephson junction, shunt capacitor, and microwave coupling capacitor are fabricated using optical lithography and electron-beam evaporation.
The bottom electrodes of the capacitors are deposited in the same step as the junction, followed by alumina evaporation constituting the first \qty{30}{\nano\metre} of the insulators.
The rest of the alumina insulation (\qty{60}{\nano\metre}) and the top electrodes (aluminum, \qty{200}{\nano\metre}) are deposited after a second round of lithography.
Electrical connection to the junction at the bottom-electrode level is made via wirebonds that break through the insulating layers.

~

\section{\label{app:calib} Calibration}

In the main text we discuss two methods for calibrating the microwave power, i.e. obtaining values $A^{'c}_n (\omega)$ of the drive amplitude that maximize a given $n$th-order Shapiro step at different frequencies.
Calibrating at discrete values $\omega_i$ separated by \qty{10} to \qty{100}{\MHz} is sufficient, and intermediate frequencies can be obtained by interpolation.

The photon-assisted tunneling (PAT) method consists in choosing a reference PAT current $I^*$ and finding the microwave drive amplitude $A^*(\omega_i)$ at which this PAT current is obtained for a given frequency $\omega_i$.
In practice, we bias the junction around a voltage $V$ just below the superconducting gap, set a microwave frequency $\omega_i$ and amplitude $A$, and modulate it at a low rate $f$ (typically \qtyrange{10}{100}{\Hz}).
Then, we proceed to a lock-in measurement of the current $I$ at frequency $f$.
This enables measurement of low PAT currents, where $I \propto A$. With such a linear relationship, it is straightforward to implement a software feedback loop that converges to the value $A^*$ corresponding to the reference PAT current $I^*$. Calibration data $A^{'c}_n (\omega_i)$ are obtained by correcting values $A^*(\omega_i)$ by a constant factor that depends on chosen step $n$ and reference PAT current $I^*$.
A limitation of this method is that $V$ should be adjusted with $\omega$ because the PAT current has a stepped distribution that shifts with drive frequency.
Furthermore, it is difficult to reliably set $V$ at lower drive frequencies because of the hysteresis in the $IV$ characteristic near the superconducting gap; however, because the PAT current is monotonic with $A$, it is a reliable method for moderate-accuracy calibration.

A more accurate calibration method uses the suppression of the supercurrent $I_0$ with drive power.
The approximate position of the first zero of $I_0 = I_c J_0 (A)$ is determined using the PAT calibration method.
For each drive frequency $\omega_i$, the amplitude $A$ is swept around the position of the predicted zero.
We record the amplitude $A_{0}^{c}(\omega_i)$ at which the junction switches to nonzero voltage.
This constitutes the calibrated reference from which, using the properties of Bessel functions, we calculate $A_{n}^{'c}(\omega_i)$, the amplitude that maximizes the Shapiro peak of order $n$.
Experimentally, the supercurrent suppression is detected by measuring $V$ while applying (via $V_b$) a small bias current $I = I_b \ll I_c$: when supercurrent is reduced to $I_0 \lesssim I_b$, the junction switches from zero-voltage to a finite voltage $R I_b$.
Here, $I_b$ is chosen to be as small as permitted by the signal-to-noise ratio of $V$.
%To compensate for any bias offset, $A_{0}^{c}$ can be determined as the average of amplitude $A_{0}^{c-}$ at which the voltage switches to $V>0$ before total supercurrent suppression and the amplitude $A_{0}^{c+}$ at which the junction retraps to $V=0$ as $I_0 > I_b$ when amplitude has increased beyond enough the Bessel node.

\bibliography{main.bib}

\begin{thebibliography}{36}%
\makeatletter
\providecommand \@ifxundefined [1]{%
 \@ifx{#1\undefined}
}%
\providecommand \@ifnum [1]{%
 \ifnum #1\expandafter \@firstoftwo
 \else \expandafter \@secondoftwo
 \fi
}%
\providecommand \@ifx [1]{%
 \ifx #1\expandafter \@firstoftwo
 \else \expandafter \@secondoftwo
 \fi
}%
\providecommand \natexlab [1]{#1}%
\providecommand \enquote  [1]{``#1''}%
\providecommand \bibnamefont  [1]{#1}%
\providecommand \bibfnamefont [1]{#1}%
\providecommand \citenamefont [1]{#1}%
\providecommand \href@noop [0]{\@secondoftwo}%
\providecommand \href [0]{\begingroup \@sanitize@url \@href}%
\providecommand \@href[1]{\@@startlink{#1}\@@href}%
\providecommand \@@href[1]{\endgroup#1\@@endlink}%
\providecommand \@sanitize@url [0]{\catcode `\\12\catcode `\$12\catcode
  `\&12\catcode `\#12\catcode `\^12\catcode `\_12\catcode `\%12\relax}%
\providecommand \@@startlink[1]{}%
\providecommand \@@endlink[0]{}%
\providecommand \url  [0]{\begingroup\@sanitize@url \@url }%
\providecommand \@url [1]{\endgroup\@href {#1}{\urlprefix }}%
\providecommand \urlprefix  [0]{URL }%
\providecommand \Eprint [0]{\href }%
\providecommand \doibase [0]{https://doi.org/}%
\providecommand \selectlanguage [0]{\@gobble}%
\providecommand \bibinfo  [0]{\@secondoftwo}%
\providecommand \bibfield  [0]{\@secondoftwo}%
\providecommand \translation [1]{[#1]}%
\providecommand \BibitemOpen [0]{}%
\providecommand \bibitemStop [0]{}%
\providecommand \bibitemNoStop [0]{.\EOS\space}%
\providecommand \EOS [0]{\spacefactor3000\relax}%
\providecommand \BibitemShut  [1]{\csname bibitem#1\endcsname}%
\let\auto@bib@innerbib\@empty
%</preamble>
\bibitem [{\citenamefont {Jeanneret}\ and\ \citenamefont
  {Benz}(2009)}]{jeanneret_application_2009}%
  \BibitemOpen
  \bibfield  {author} {\bibinfo {author} {\bibfnamefont {B.}~\bibnamefont
  {Jeanneret}}\ and\ \bibinfo {author} {\bibfnamefont {S.~P.}\ \bibnamefont
  {Benz}},\ }\bibfield  {title} {\bibinfo {title} {Application of the
  {Josephson} effect in electrical metrology},\ }\href
  {https://doi.org/10.1140/epjst/e2009-01050-6} {\bibfield  {journal} {\bibinfo
   {journal} {The European Physical Journal Special Topics}\ }\textbf {\bibinfo
  {volume} {172}},\ \bibinfo {pages} {181} (\bibinfo {year}
  {2009})}\BibitemShut {NoStop}%
\bibitem [{\citenamefont {Mohr}\ and\ \citenamefont
  {Taylor}(2002)}]{mohr_codata_2002}%
  \BibitemOpen
  \bibfield  {author} {\bibinfo {author} {\bibfnamefont {P.~J.}\ \bibnamefont
  {Mohr}}\ and\ \bibinfo {author} {\bibfnamefont {B.~N.}\ \bibnamefont
  {Taylor}},\ }\bibfield  {title} {\bibinfo {title} {{CODATA} recommended
  values of the fundamental physical constants: 2002},\ }\href@noop {}
  {\bibfield  {journal} {\bibinfo  {journal} {Rev. Mod. Phys.}\ }\textbf
  {\bibinfo {volume} {77}},\ \bibinfo {pages} {107} (\bibinfo {year}
  {2002})}\BibitemShut {NoStop}%
\bibitem [{\citenamefont {Rüfenacht}\ \emph {et~al.}(2018)\citenamefont
  {Rüfenacht}, \citenamefont {Flowers-Jacobs},\ and\ \citenamefont
  {Benz}}]{Rufenacht_Impact_2018}%
  \BibitemOpen
  \bibfield  {author} {\bibinfo {author} {\bibfnamefont {A.}~\bibnamefont
  {Rüfenacht}}, \bibinfo {author} {\bibfnamefont {N.~E.}\ \bibnamefont
  {Flowers-Jacobs}},\ and\ \bibinfo {author} {\bibfnamefont {S.~P.}\
  \bibnamefont {Benz}},\ }\bibfield  {title} {\bibinfo {title} {Impact of the
  latest generation of {Josephson} voltage standards in ac and dc electric
  metrology},\ }\href {https://doi.org/10.1088/1681-7575/aad41a} {\bibfield
  {journal} {\bibinfo  {journal} {Metrologia}\ }\textbf {\bibinfo {volume}
  {55}},\ \bibinfo {pages} {S152} (\bibinfo {year} {2018})}\BibitemShut
  {NoStop}%
\bibitem [{\citenamefont {Burroughs}\ \emph {et~al.}(2011)\citenamefont
  {Burroughs}, \citenamefont {Dresselhaus}, \citenamefont {Rufenacht},
  \citenamefont {Olaya}, \citenamefont {Elsbury}, \citenamefont {Tang},\ and\
  \citenamefont {Benz}}]{burroughs_nist_2011}%
  \BibitemOpen
  \bibfield  {author} {\bibinfo {author} {\bibfnamefont {C.~J.}\ \bibnamefont
  {Burroughs}}, \bibinfo {author} {\bibfnamefont {P.~D.}\ \bibnamefont
  {Dresselhaus}}, \bibinfo {author} {\bibfnamefont {A.}~\bibnamefont
  {Rufenacht}}, \bibinfo {author} {\bibfnamefont {D.}~\bibnamefont {Olaya}},
  \bibinfo {author} {\bibfnamefont {M.~M.}\ \bibnamefont {Elsbury}}, \bibinfo
  {author} {\bibfnamefont {Y.-H.}\ \bibnamefont {Tang}},\ and\ \bibinfo
  {author} {\bibfnamefont {S.~P.}\ \bibnamefont {Benz}},\ }\bibfield  {title}
  {\bibinfo {title} {{NIST} 10 {V} {Programmable} {Josephson} {Voltage}
  {Standard} {System}},\ }\href {https://doi.org/10.1109/TIM.2010.2101191}
  {\bibfield  {journal} {\bibinfo  {journal} {IEEE Transactions on
  Instrumentation and Measurement}\ }\textbf {\bibinfo {volume} {60}},\
  \bibinfo {pages} {2482} (\bibinfo {year} {2011})}\BibitemShut {NoStop}%
\bibitem [{\citenamefont {Benz}\ and\ \citenamefont
  {Hamilton}(1996)}]{benz_pulsedriven_1996}%
  \BibitemOpen
  \bibfield  {author} {\bibinfo {author} {\bibfnamefont {S.~P.}\ \bibnamefont
  {Benz}}\ and\ \bibinfo {author} {\bibfnamefont {C.~A.}\ \bibnamefont
  {Hamilton}},\ }\bibfield  {title} {\bibinfo {title} {A pulse‐driven
  programmable {Josephson} voltage standard},\ }\href
  {https://doi.org/10.1063/1.115814} {\bibfield  {journal} {\bibinfo  {journal}
  {Applied Physics Letters}\ }\textbf {\bibinfo {volume} {68}},\ \bibinfo
  {pages} {3171} (\bibinfo {year} {1996})}\BibitemShut {NoStop}%
\bibitem [{\citenamefont {Girit}\ and\ \citenamefont
  {Smirr}(2023)}]{giritsmirr2023}%
  \BibitemOpen
  \bibfield  {author} {\bibinfo {author} {\bibfnamefont {C.}~\bibnamefont
  {Girit}}\ and\ \bibinfo {author} {\bibfnamefont {J.-L.}\ \bibnamefont
  {Smirr}},\ }\href@noop {} {\bibinfo {title} {Voltage source and method for
  calibrating this voltage source}} (\bibinfo {year} {2023}),\ \bibinfo {note}
  {{}FR patent FR3114171B1, US patent US20230341880A1 pend.}\BibitemShut
  {Stop}%
\bibitem [{\citenamefont {Griesmar}\ \emph {et~al.}(2021)\citenamefont
  {Griesmar}, \citenamefont {Rodriguez}, \citenamefont {Benzoni}, \citenamefont
  {Pillet}, \citenamefont {Smirr}, \citenamefont {Lafont},\ and\ \citenamefont
  {Girit}}]{griesmar_superconducting_2021}%
  \BibitemOpen
  \bibfield  {author} {\bibinfo {author} {\bibfnamefont {J.}~\bibnamefont
  {Griesmar}}, \bibinfo {author} {\bibfnamefont {R.~H.}\ \bibnamefont
  {Rodriguez}}, \bibinfo {author} {\bibfnamefont {V.}~\bibnamefont {Benzoni}},
  \bibinfo {author} {\bibfnamefont {J.-D.}\ \bibnamefont {Pillet}}, \bibinfo
  {author} {\bibfnamefont {J.-L.}\ \bibnamefont {Smirr}}, \bibinfo {author}
  {\bibfnamefont {F.}~\bibnamefont {Lafont}},\ and\ \bibinfo {author}
  {\bibfnamefont {{\c{C}}.~{\"{O}}.}\ \bibnamefont {Girit}},\ }\bibfield
  {title} {\bibinfo {title} {Superconducting on-chip spectrometer for
  mesoscopic quantum systems},\ }\href
  {https://doi.org/10.1103/PhysRevResearch.3.043078} {\bibfield  {journal}
  {\bibinfo  {journal} {Physical Review Research}\ }\textbf {\bibinfo {volume}
  {3}},\ \bibinfo {pages} {043078} (\bibinfo {year} {2021})}\BibitemShut
  {NoStop}%
\bibitem [{\citenamefont {Peyruchat}\ \emph {et~al.}(2021)\citenamefont
  {Peyruchat}, \citenamefont {Griesmar}, \citenamefont {Pillet},\ and\
  \citenamefont {Girit}}]{peyruchat_transconductance_2021}%
  \BibitemOpen
  \bibfield  {author} {\bibinfo {author} {\bibfnamefont {L.}~\bibnamefont
  {Peyruchat}}, \bibinfo {author} {\bibfnamefont {J.}~\bibnamefont {Griesmar}},
  \bibinfo {author} {\bibfnamefont {J.-D.}\ \bibnamefont {Pillet}},\ and\
  \bibinfo {author} {\bibfnamefont {{\c{C}}.~{\"{O}}.}\ \bibnamefont {Girit}},\
  }\bibfield  {title} {\bibinfo {title} {Transconductance quantization in a
  topological {Josephson} tunnel junction circuit},\ }\href
  {https://doi.org/10.1103/PhysRevResearch.3.013289} {\bibfield  {journal}
  {\bibinfo  {journal} {Physical Review Research}\ }\textbf {\bibinfo {volume}
  {3}},\ \bibinfo {pages} {013289} (\bibinfo {year} {2021})}\BibitemShut
  {NoStop}%
\bibitem [{\citenamefont {Jebari}\ \emph {et~al.}(2018)\citenamefont {Jebari},
  \citenamefont {Blanchet}, \citenamefont {Grimm}, \citenamefont {Hazra},
  \citenamefont {Albert}, \citenamefont {Joyez}, \citenamefont {Vion},
  \citenamefont {Estève}, \citenamefont {Portier},\ and\ \citenamefont
  {Hofheinz}}]{jebari_near-quantum-limited_2018}%
  \BibitemOpen
  \bibfield  {author} {\bibinfo {author} {\bibfnamefont {S.}~\bibnamefont
  {Jebari}}, \bibinfo {author} {\bibfnamefont {F.}~\bibnamefont {Blanchet}},
  \bibinfo {author} {\bibfnamefont {A.}~\bibnamefont {Grimm}}, \bibinfo
  {author} {\bibfnamefont {D.}~\bibnamefont {Hazra}}, \bibinfo {author}
  {\bibfnamefont {R.}~\bibnamefont {Albert}}, \bibinfo {author} {\bibfnamefont
  {P.}~\bibnamefont {Joyez}}, \bibinfo {author} {\bibfnamefont
  {D.}~\bibnamefont {Vion}}, \bibinfo {author} {\bibfnamefont {D.}~\bibnamefont
  {Estève}}, \bibinfo {author} {\bibfnamefont {F.}~\bibnamefont {Portier}},\
  and\ \bibinfo {author} {\bibfnamefont {M.}~\bibnamefont {Hofheinz}},\
  }\bibfield  {title} {\bibinfo {title} {Near-quantum-limited amplification
  from inelastic {Cooper}-pair tunnelling},\ }\href
  {https://doi.org/10.1038/s41928-018-0055-7} {\bibfield  {journal} {\bibinfo
  {journal} {Nature Electronics}\ }\textbf {\bibinfo {volume} {1}},\ \bibinfo
  {pages} {223} (\bibinfo {year} {2018})}\BibitemShut {NoStop}%
\bibitem [{\citenamefont {Mendes}\ \emph {et~al.}(2019)\citenamefont {Mendes},
  \citenamefont {Jezouin}, \citenamefont {Joyez}, \citenamefont {Reulet},
  \citenamefont {Blais}, \citenamefont {Portier}, \citenamefont {Mora},\ and\
  \citenamefont {Altimiras}}]{mendes_parametric_2019}%
  \BibitemOpen
  \bibfield  {author} {\bibinfo {author} {\bibfnamefont {U.~C.}\ \bibnamefont
  {Mendes}}, \bibinfo {author} {\bibfnamefont {S.}~\bibnamefont {Jezouin}},
  \bibinfo {author} {\bibfnamefont {P.}~\bibnamefont {Joyez}}, \bibinfo
  {author} {\bibfnamefont {B.}~\bibnamefont {Reulet}}, \bibinfo {author}
  {\bibfnamefont {A.}~\bibnamefont {Blais}}, \bibinfo {author} {\bibfnamefont
  {F.}~\bibnamefont {Portier}}, \bibinfo {author} {\bibfnamefont
  {C.}~\bibnamefont {Mora}},\ and\ \bibinfo {author} {\bibfnamefont
  {C.}~\bibnamefont {Altimiras}},\ }\bibfield  {title} {\bibinfo {title}
  {Parametric amplification and squeezing with an ac- and dc-voltage biased
  superconducting junction},\ }\href
  {https://doi.org/10.1103/PhysRevApplied.11.034035} {\bibfield  {journal}
  {\bibinfo  {journal} {Physical Review Applied}\ }\textbf {\bibinfo {volume}
  {11}},\ \bibinfo {pages} {034035} (\bibinfo {year} {2019})}\BibitemShut
  {NoStop}%
\bibitem [{\citenamefont {Grimm}\ \emph {et~al.}(2019)\citenamefont {Grimm},
  \citenamefont {Blanchet}, \citenamefont {Albert}, \citenamefont
  {Leppäkangas}, \citenamefont {Jebari}, \citenamefont {Hazra}, \citenamefont
  {Gustavo}, \citenamefont {Thomassin}, \citenamefont {Dupont-Ferrier},
  \citenamefont {Portier},\ and\ \citenamefont {Hofheinz}}]{grimm_bright_2019}%
  \BibitemOpen
  \bibfield  {author} {\bibinfo {author} {\bibfnamefont {A.}~\bibnamefont
  {Grimm}}, \bibinfo {author} {\bibfnamefont {F.}~\bibnamefont {Blanchet}},
  \bibinfo {author} {\bibfnamefont {R.}~\bibnamefont {Albert}}, \bibinfo
  {author} {\bibfnamefont {J.}~\bibnamefont {Leppäkangas}}, \bibinfo {author}
  {\bibfnamefont {S.}~\bibnamefont {Jebari}}, \bibinfo {author} {\bibfnamefont
  {D.}~\bibnamefont {Hazra}}, \bibinfo {author} {\bibfnamefont
  {F.}~\bibnamefont {Gustavo}}, \bibinfo {author} {\bibfnamefont {J.-L.}\
  \bibnamefont {Thomassin}}, \bibinfo {author} {\bibfnamefont {E.}~\bibnamefont
  {Dupont-Ferrier}}, \bibinfo {author} {\bibfnamefont {F.}~\bibnamefont
  {Portier}},\ and\ \bibinfo {author} {\bibfnamefont {M.}~\bibnamefont
  {Hofheinz}},\ }\bibfield  {title} {\bibinfo {title} {Bright {On}-{Demand}
  {Source} of {Antibunched} {Microwave} {Photons} {Based} on {Inelastic}
  {Cooper} {Pair} {Tunneling}},\ }\href
  {https://doi.org/10.1103/PhysRevX.9.021016} {\bibfield  {journal} {\bibinfo
  {journal} {Physical Review X}\ }\textbf {\bibinfo {volume} {9}},\ \bibinfo
  {pages} {021016} (\bibinfo {year} {2019})}\BibitemShut {NoStop}%
\bibitem [{\citenamefont {Rolland}\ \emph {et~al.}(2019)\citenamefont
  {Rolland}, \citenamefont {Peugeot}, \citenamefont {Dambach}, \citenamefont
  {Westig}, \citenamefont {Kubala}, \citenamefont {Mukharsky}, \citenamefont
  {Altimiras}, \citenamefont {le~Sueur}, \citenamefont {Joyez}, \citenamefont
  {Vion}, \citenamefont {Roche}, \citenamefont {Esteve}, \citenamefont
  {Ankerhold},\ and\ \citenamefont {Portier}}]{rolland_antibunched_2019}%
  \BibitemOpen
  \bibfield  {author} {\bibinfo {author} {\bibfnamefont {C.}~\bibnamefont
  {Rolland}}, \bibinfo {author} {\bibfnamefont {A.}~\bibnamefont {Peugeot}},
  \bibinfo {author} {\bibfnamefont {S.}~\bibnamefont {Dambach}}, \bibinfo
  {author} {\bibfnamefont {M.}~\bibnamefont {Westig}}, \bibinfo {author}
  {\bibfnamefont {B.}~\bibnamefont {Kubala}}, \bibinfo {author} {\bibfnamefont
  {Y.}~\bibnamefont {Mukharsky}}, \bibinfo {author} {\bibfnamefont
  {C.}~\bibnamefont {Altimiras}}, \bibinfo {author} {\bibfnamefont
  {H.}~\bibnamefont {le~Sueur}}, \bibinfo {author} {\bibfnamefont
  {P.}~\bibnamefont {Joyez}}, \bibinfo {author} {\bibfnamefont
  {D.}~\bibnamefont {Vion}}, \bibinfo {author} {\bibfnamefont {P.}~\bibnamefont
  {Roche}}, \bibinfo {author} {\bibfnamefont {D.}~\bibnamefont {Esteve}},
  \bibinfo {author} {\bibfnamefont {J.}~\bibnamefont {Ankerhold}},\ and\
  \bibinfo {author} {\bibfnamefont {F.}~\bibnamefont {Portier}},\ }\bibfield
  {title} {\bibinfo {title} {Antibunched {Photons} {Emitted} by a dc-{Biased}
  {Josephson} {Junction}},\ }\href
  {https://doi.org/10.1103/PhysRevLett.122.186804} {\bibfield  {journal}
  {\bibinfo  {journal} {Physical Review Letters}\ }\textbf {\bibinfo {volume}
  {122}},\ \bibinfo {pages} {186804} (\bibinfo {year} {2019})}\BibitemShut
  {NoStop}%
\bibitem [{\citenamefont {Peugeot}\ \emph {et~al.}(2021)\citenamefont
  {Peugeot}, \citenamefont {Ménard}, \citenamefont {Dambach}, \citenamefont
  {Westig}, \citenamefont {Kubala}, \citenamefont {Mukharsky}, \citenamefont
  {Altimiras}, \citenamefont {Joyez}, \citenamefont {Vion}, \citenamefont
  {Roche}, \citenamefont {Esteve}, \citenamefont {Milman}, \citenamefont
  {Leppäkangas}, \citenamefont {Johansson}, \citenamefont {Hofheinz},
  \citenamefont {Ankerhold},\ and\ \citenamefont
  {Portier}}]{peugeot_generating_2021}%
  \BibitemOpen
  \bibfield  {author} {\bibinfo {author} {\bibfnamefont {A.}~\bibnamefont
  {Peugeot}}, \bibinfo {author} {\bibfnamefont {G.}~\bibnamefont {Ménard}},
  \bibinfo {author} {\bibfnamefont {S.}~\bibnamefont {Dambach}}, \bibinfo
  {author} {\bibfnamefont {M.}~\bibnamefont {Westig}}, \bibinfo {author}
  {\bibfnamefont {B.}~\bibnamefont {Kubala}}, \bibinfo {author} {\bibfnamefont
  {Y.}~\bibnamefont {Mukharsky}}, \bibinfo {author} {\bibfnamefont
  {C.}~\bibnamefont {Altimiras}}, \bibinfo {author} {\bibfnamefont
  {P.}~\bibnamefont {Joyez}}, \bibinfo {author} {\bibfnamefont
  {D.}~\bibnamefont {Vion}}, \bibinfo {author} {\bibfnamefont {P.}~\bibnamefont
  {Roche}}, \bibinfo {author} {\bibfnamefont {D.}~\bibnamefont {Esteve}},
  \bibinfo {author} {\bibfnamefont {P.}~\bibnamefont {Milman}}, \bibinfo
  {author} {\bibfnamefont {J.}~\bibnamefont {Leppäkangas}}, \bibinfo {author}
  {\bibfnamefont {G.}~\bibnamefont {Johansson}}, \bibinfo {author}
  {\bibfnamefont {M.}~\bibnamefont {Hofheinz}}, \bibinfo {author}
  {\bibfnamefont {J.}~\bibnamefont {Ankerhold}},\ and\ \bibinfo {author}
  {\bibfnamefont {F.}~\bibnamefont {Portier}},\ }\bibfield  {title} {\bibinfo
  {title} {Generating {Two} {Continuous} {Entangled} {Microwave} {Beams}
  {Using} a dc-{Biased} {Josephson} {Junction}},\ }\href
  {https://doi.org/10.1103/PhysRevX.11.031008} {\bibfield  {journal} {\bibinfo
  {journal} {Physical Review X}\ }\textbf {\bibinfo {volume} {11}},\ \bibinfo
  {pages} {031008} (\bibinfo {year} {2021})}\BibitemShut {NoStop}%
\bibitem [{\citenamefont {Ma}\ \emph {et~al.}(2021)\citenamefont {Ma},
  \citenamefont {Li}, \citenamefont {Ren}, \citenamefont {Xie},\ and\
  \citenamefont {Li}}]{ma_antibunched_2021}%
  \BibitemOpen
  \bibfield  {author} {\bibinfo {author} {\bibfnamefont {S.-l.}\ \bibnamefont
  {Ma}}, \bibinfo {author} {\bibfnamefont {X.-K.}\ \bibnamefont {Li}}, \bibinfo
  {author} {\bibfnamefont {Y.-L.}\ \bibnamefont {Ren}}, \bibinfo {author}
  {\bibfnamefont {J.-K.}\ \bibnamefont {Xie}},\ and\ \bibinfo {author}
  {\bibfnamefont {F.-L.}\ \bibnamefont {Li}},\ }\bibfield  {title} {\bibinfo
  {title} {Antibunched $n$-photon bundles emitted by a {Josephson} photonic
  device},\ }\href {https://doi.org/10.1103/PhysRevResearch.3.043020}
  {\bibfield  {journal} {\bibinfo  {journal} {Physical Review Research}\
  }\textbf {\bibinfo {volume} {3}},\ \bibinfo {pages} {043020} (\bibinfo {year}
  {2021})}\BibitemShut {NoStop}%
\bibitem [{\citenamefont {Lörch}\ \emph {et~al.}(2018)\citenamefont {Lörch},
  \citenamefont {Bruder}, \citenamefont {Brunner},\ and\ \citenamefont
  {Hofer}}]{lorch_optimal_2018}%
  \BibitemOpen
  \bibfield  {author} {\bibinfo {author} {\bibfnamefont {N.}~\bibnamefont
  {Lörch}}, \bibinfo {author} {\bibfnamefont {C.}~\bibnamefont {Bruder}},
  \bibinfo {author} {\bibfnamefont {N.}~\bibnamefont {Brunner}},\ and\ \bibinfo
  {author} {\bibfnamefont {P.~P.}\ \bibnamefont {Hofer}},\ }\bibfield  {title}
  {\bibinfo {title} {Optimal work extraction from quantum states by
  photo-assisted {Cooper} pair tunneling},\ }\href
  {https://doi.org/10.1088/2058-9565/aacbf3} {\bibfield  {journal} {\bibinfo
  {journal} {Quantum Science and Technology}\ }\textbf {\bibinfo {volume}
  {3}},\ \bibinfo {pages} {035014} (\bibinfo {year} {2018})}\BibitemShut
  {NoStop}%
\bibitem [{\citenamefont {Kautz}(1996)}]{kautz_noise_1996}%
  \BibitemOpen
  \bibfield  {author} {\bibinfo {author} {\bibfnamefont {R.~L.}\ \bibnamefont
  {Kautz}},\ }\bibfield  {title} {\bibinfo {title} {Noise, chaos, and the
  {Josephson} voltage standard},\ }\href
  {https://doi.org/10.1088/0034-4885/59/8/001} {\bibfield  {journal} {\bibinfo
  {journal} {Reports on Progress in Physics}\ }\textbf {\bibinfo {volume}
  {59}},\ \bibinfo {pages} {935} (\bibinfo {year} {1996})}\BibitemShut
  {NoStop}%
\bibitem [{\citenamefont {Orlando}\ and\ \citenamefont
  {Delin}(1991)}]{orlandoFoundationsAppliedSuperconductivity1991}%
  \BibitemOpen
  \bibfield  {author} {\bibinfo {author} {\bibfnamefont {T.}~\bibnamefont
  {Orlando}}\ and\ \bibinfo {author} {\bibfnamefont {K.~A.}\ \bibnamefont
  {Delin}},\ }\href@noop {} {\emph {\bibinfo {title} {Foundations of {{Applied
  Superconductivity}}}}}\ (\bibinfo  {publisher} {Addison-Wesley},\ \bibinfo
  {address} {Reading, Mass.},\ \bibinfo {year} {1991})\BibitemShut {NoStop}%
\bibitem [{\citenamefont {Shapiro}\ \emph {et~al.}(1964)\citenamefont
  {Shapiro}, \citenamefont {Janus},\ and\ \citenamefont
  {Holly}}]{shapiro_effect_1964}%
  \BibitemOpen
  \bibfield  {author} {\bibinfo {author} {\bibfnamefont {S.}~\bibnamefont
  {Shapiro}}, \bibinfo {author} {\bibfnamefont {A.~R.}\ \bibnamefont {Janus}},\
  and\ \bibinfo {author} {\bibfnamefont {S.}~\bibnamefont {Holly}},\ }\bibfield
   {title} {\bibinfo {title} {Effect of {Microwaves} on {Josephson} {Currents}
  in {Superconducting} {Tunneling}},\ }\href
  {https://doi.org/10.1103/RevModPhys.36.223} {\bibfield  {journal} {\bibinfo
  {journal} {Reviews of Modern Physics}\ }\textbf {\bibinfo {volume} {36}},\
  \bibinfo {pages} {223} (\bibinfo {year} {1964})}\BibitemShut {NoStop}%
\bibitem [{\citenamefont {{Likharev}}(1986)}]{likharev_dynamics_1986}%
  \BibitemOpen
  \bibfield  {author} {\bibinfo {author} {\bibnamefont {{Likharev}}},\
  }\href@noop {} {\emph {\bibinfo {title} {Dynamics of {Josephson} {Junctions}
  and {Circuits}}}}\ (\bibinfo  {publisher} {CRC Press},\ \bibinfo {year}
  {1986})\BibitemShut {NoStop}%
\bibitem [{\citenamefont {Richards}\ \emph {et~al.}(1973)\citenamefont
  {Richards}, \citenamefont {Auracher},\ and\ \citenamefont
  {Van~Duzer}}]{richards_millimeter_1973}%
  \BibitemOpen
  \bibfield  {author} {\bibinfo {author} {\bibfnamefont {P.}~\bibnamefont
  {Richards}}, \bibinfo {author} {\bibfnamefont {F.}~\bibnamefont {Auracher}},\
  and\ \bibinfo {author} {\bibfnamefont {T.}~\bibnamefont {Van~Duzer}},\
  }\bibfield  {title} {\bibinfo {title} {Millimeter and submillimeter wave
  detection and mixing with superconducting weak links},\ }\href
  {https://doi.org/10.1109/PROC.1973.8967} {\bibfield  {journal} {\bibinfo
  {journal} {Proceedings of the IEEE}\ }\textbf {\bibinfo {volume} {61}},\
  \bibinfo {pages} {36} (\bibinfo {year} {1973})}\BibitemShut {NoStop}%
\bibitem [{\citenamefont {Zmuidzinas}\ and\ \citenamefont
  {Richards}(2004)}]{zmuidzinas_superconducting_2004}%
  \BibitemOpen
  \bibfield  {author} {\bibinfo {author} {\bibfnamefont {J.}~\bibnamefont
  {Zmuidzinas}}\ and\ \bibinfo {author} {\bibfnamefont {P.}~\bibnamefont
  {Richards}},\ }\bibfield  {title} {\bibinfo {title} {Superconducting
  detectors and mixers for millimeter and submillimeter astrophysics},\ }\href
  {https://doi.org/10.1109/JPROC.2004.833670} {\bibfield  {journal} {\bibinfo
  {journal} {Proceedings of the IEEE}\ }\textbf {\bibinfo {volume} {92}},\
  \bibinfo {pages} {1597} (\bibinfo {year} {2004})}\BibitemShut {NoStop}%
\bibitem [{\citenamefont {Tucker}\ and\ \citenamefont
  {Feldman}(1985)}]{tucker_quantum_1985}%
  \BibitemOpen
  \bibfield  {author} {\bibinfo {author} {\bibfnamefont {J.~R.}\ \bibnamefont
  {Tucker}}\ and\ \bibinfo {author} {\bibfnamefont {M.~J.}\ \bibnamefont
  {Feldman}},\ }\bibfield  {title} {\bibinfo {title} {Quantum detection at
  millimeter wavelengths},\ }\href {https://doi.org/10.1103/RevModPhys.57.1055}
  {\bibfield  {journal} {\bibinfo  {journal} {Reviews of Modern Physics}\
  }\textbf {\bibinfo {volume} {57}},\ \bibinfo {pages} {1055} (\bibinfo {year}
  {1985})}\BibitemShut {NoStop}%
\bibitem [{\citenamefont {Levinsen}\ \emph {et~al.}(1977)\citenamefont
  {Levinsen}, \citenamefont {Chiao}, \citenamefont {Feldman},\ and\
  \citenamefont {Tucker}}]{levinsen_inverse_1977}%
  \BibitemOpen
  \bibfield  {author} {\bibinfo {author} {\bibfnamefont {M.~T.}\ \bibnamefont
  {Levinsen}}, \bibinfo {author} {\bibfnamefont {R.~Y.}\ \bibnamefont {Chiao}},
  \bibinfo {author} {\bibfnamefont {M.~J.}\ \bibnamefont {Feldman}},\ and\
  \bibinfo {author} {\bibfnamefont {B.~A.}\ \bibnamefont {Tucker}},\ }\bibfield
   {title} {\bibinfo {title} {An inverse ac {Josephson} effect voltage
  standard},\ }\href {https://doi.org/10.1063/1.89520} {\bibfield  {journal}
  {\bibinfo  {journal} {Applied Physics Letters}\ }\textbf {\bibinfo {volume}
  {31}},\ \bibinfo {pages} {776} (\bibinfo {year} {1977})}\BibitemShut
  {NoStop}%
\bibitem [{\citenamefont {Kautz}(1980)}]{kautz_proposed_1980}%
  \BibitemOpen
  \bibfield  {author} {\bibinfo {author} {\bibfnamefont {R.~L.}\ \bibnamefont
  {Kautz}},\ }\bibfield  {title} {\bibinfo {title} {On a proposed
  {Josephson}‐effect voltage standard at zero current bias},\ }\href
  {https://doi.org/10.1063/1.91497} {\bibfield  {journal} {\bibinfo  {journal}
  {Applied Physics Letters}\ }\textbf {\bibinfo {volume} {36}},\ \bibinfo
  {pages} {386} (\bibinfo {year} {1980})}\BibitemShut {NoStop}%
\bibitem [{\citenamefont {Yanson}\ \emph {et~al.}(1965)\citenamefont {Yanson},
  \citenamefont {Svistunov},\ and\ \citenamefont
  {Dmitrenko}}]{yanson_experimental_1965}%
  \BibitemOpen
  \bibfield  {author} {\bibinfo {author} {\bibfnamefont {I.}~\bibnamefont
  {Yanson}}, \bibinfo {author} {\bibfnamefont {V.}~\bibnamefont {Svistunov}},\
  and\ \bibinfo {author} {\bibfnamefont {I.}~\bibnamefont {Dmitrenko}},\
  }\bibfield  {title} {\bibinfo {title} {Experimental observation of the tunnel
  effect for {Cooper} pairs with the emission of photons},\ }\href@noop {}
  {\bibfield  {journal} {\bibinfo  {journal} {Sov. Phys. JETP}\ }\textbf
  {\bibinfo {volume} {21}},\ \bibinfo {pages} {650} (\bibinfo {year}
  {1965})}\BibitemShut {NoStop}%
\bibitem [{\citenamefont {Hofheinz}\ \emph {et~al.}(2011)\citenamefont
  {Hofheinz}, \citenamefont {Portier}, \citenamefont {Baudouin}, \citenamefont
  {Joyez}, \citenamefont {Vion}, \citenamefont {Bertet}, \citenamefont
  {Roche},\ and\ \citenamefont {Esteve}}]{hofheinz_bright_2011}%
  \BibitemOpen
  \bibfield  {author} {\bibinfo {author} {\bibfnamefont {M.}~\bibnamefont
  {Hofheinz}}, \bibinfo {author} {\bibfnamefont {F.}~\bibnamefont {Portier}},
  \bibinfo {author} {\bibfnamefont {Q.}~\bibnamefont {Baudouin}}, \bibinfo
  {author} {\bibfnamefont {P.}~\bibnamefont {Joyez}}, \bibinfo {author}
  {\bibfnamefont {D.}~\bibnamefont {Vion}}, \bibinfo {author} {\bibfnamefont
  {P.}~\bibnamefont {Bertet}}, \bibinfo {author} {\bibfnamefont
  {P.}~\bibnamefont {Roche}},\ and\ \bibinfo {author} {\bibfnamefont
  {D.}~\bibnamefont {Esteve}},\ }\bibfield  {title} {\bibinfo {title} {Bright
  {Side} of the {Coulomb} {Blockade}},\ }\href
  {https://doi.org/10.1103/PhysRevLett.106.217005} {\bibfield  {journal}
  {\bibinfo  {journal} {Physical Review Letters}\ }\textbf {\bibinfo {volume}
  {106}},\ \bibinfo {pages} {217005} (\bibinfo {year} {2011})}\BibitemShut
  {NoStop}%
\bibitem [{\citenamefont {Cassidy}\ \emph {et~al.}(2017)\citenamefont
  {Cassidy}, \citenamefont {Bruno}, \citenamefont {Rubbert}, \citenamefont
  {Irfan}, \citenamefont {Kammhuber}, \citenamefont {Schouten}, \citenamefont
  {Akhmerov},\ and\ \citenamefont {Kouwenhoven}}]{cassidy_demonstration_2017}%
  \BibitemOpen
  \bibfield  {author} {\bibinfo {author} {\bibfnamefont {M.~C.}\ \bibnamefont
  {Cassidy}}, \bibinfo {author} {\bibfnamefont {A.}~\bibnamefont {Bruno}},
  \bibinfo {author} {\bibfnamefont {S.}~\bibnamefont {Rubbert}}, \bibinfo
  {author} {\bibfnamefont {M.}~\bibnamefont {Irfan}}, \bibinfo {author}
  {\bibfnamefont {J.}~\bibnamefont {Kammhuber}}, \bibinfo {author}
  {\bibfnamefont {R.~N.}\ \bibnamefont {Schouten}}, \bibinfo {author}
  {\bibfnamefont {A.~R.}\ \bibnamefont {Akhmerov}},\ and\ \bibinfo {author}
  {\bibfnamefont {L.~P.}\ \bibnamefont {Kouwenhoven}},\ }\bibfield  {title}
  {\bibinfo {title} {Demonstration of an ac {Josephson} junction laser},\
  }\href {https://doi.org/10.1126/science.aah6640} {\bibfield  {journal}
  {\bibinfo  {journal} {Science}\ }\textbf {\bibinfo {volume} {355}},\ \bibinfo
  {pages} {939} (\bibinfo {year} {2017})}\BibitemShut {NoStop}%
\bibitem [{\citenamefont {Ingold}\ and\ \citenamefont
  {Nazarov}(1992)}]{ingold_1992}%
  \BibitemOpen
  \bibfield  {author} {\bibinfo {author} {\bibfnamefont {G.-L.}\ \bibnamefont
  {Ingold}}\ and\ \bibinfo {author} {\bibfnamefont {Y.~V.}\ \bibnamefont
  {Nazarov}},\ }\bibinfo {title} {Charge tunneling rates in ultrasmall
  junctions},\ in\ \href {https://doi.org/10.1007/978-1-4757-2166-9_2} {\emph
  {\bibinfo {booktitle} {Single Charge Tunneling: Coulomb Blockade Phenomena In
  Nanostructures}}}\ (\bibinfo  {publisher} {Springer US},\ \bibinfo {address}
  {Boston, MA},\ \bibinfo {year} {1992})\ Chap.~\bibinfo {chapter} {2}, pp.\
  \bibinfo {pages} {21--107}\BibitemShut {NoStop}%
\bibitem [{\citenamefont {Holst}\ \emph {et~al.}(1994)\citenamefont {Holst},
  \citenamefont {Esteve}, \citenamefont {Urbina},\ and\ \citenamefont
  {Devoret}}]{holst_1994}%
  \BibitemOpen
  \bibfield  {author} {\bibinfo {author} {\bibfnamefont {T.}~\bibnamefont
  {Holst}}, \bibinfo {author} {\bibfnamefont {D.}~\bibnamefont {Esteve}},
  \bibinfo {author} {\bibfnamefont {C.}~\bibnamefont {Urbina}},\ and\ \bibinfo
  {author} {\bibfnamefont {M.~H.}\ \bibnamefont {Devoret}},\ }\bibfield
  {title} {\bibinfo {title} {Effect of a transmission line resonator on a small
  capacitance tunnel junction},\ }\href
  {https://doi.org/10.1103/PhysRevLett.73.3455} {\bibfield  {journal} {\bibinfo
   {journal} {Phys. Rev. Lett.}\ }\textbf {\bibinfo {volume} {73}},\ \bibinfo
  {pages} {3455} (\bibinfo {year} {1994})}\BibitemShut {NoStop}%
\bibitem [{\citenamefont {Stewart}(1954)}]{stewart_power_1954}%
  \BibitemOpen
  \bibfield  {author} {\bibinfo {author} {\bibfnamefont {J.~L.}\ \bibnamefont
  {Stewart}},\ }\bibfield  {title} {\bibinfo {title} {The {Power} {Spectrum} of
  a {Carrier} {Frequency} {Modulated} by {Gaussian} {Noise}},\ }\href
  {https://doi.org/10.1109/JRPROC.1954.274758} {\bibfield  {journal} {\bibinfo
  {journal} {Proceedings of the IRE}\ }\textbf {\bibinfo {volume} {42}},\
  \bibinfo {pages} {1539} (\bibinfo {year} {1954})}\BibitemShut {NoStop}%
\bibitem [{\citenamefont {Peugeot}\ \emph {et~al.}(2024)\citenamefont
  {Peugeot}, \citenamefont {Riechert}, \citenamefont {Annabi}, \citenamefont
  {Balembois}, \citenamefont {Villiers}, \citenamefont {Flurin}, \citenamefont
  {Griesmar}, \citenamefont {Arrighi}, \citenamefont {Pillet},\ and\
  \citenamefont {Bretheau}}]{peugeotTwotoneSpectroscopyHighfrequency2024d}%
  \BibitemOpen
  \bibfield  {author} {\bibinfo {author} {\bibfnamefont {A.}~\bibnamefont
  {Peugeot}}, \bibinfo {author} {\bibfnamefont {H.}~\bibnamefont {Riechert}},
  \bibinfo {author} {\bibfnamefont {S.}~\bibnamefont {Annabi}}, \bibinfo
  {author} {\bibfnamefont {L.}~\bibnamefont {Balembois}}, \bibinfo {author}
  {\bibfnamefont {M.}~\bibnamefont {Villiers}}, \bibinfo {author}
  {\bibfnamefont {E.}~\bibnamefont {Flurin}}, \bibinfo {author} {\bibfnamefont
  {J.}~\bibnamefont {Griesmar}}, \bibinfo {author} {\bibfnamefont
  {E.}~\bibnamefont {Arrighi}}, \bibinfo {author} {\bibfnamefont {J.-D.}\
  \bibnamefont {Pillet}},\ and\ \bibinfo {author} {\bibfnamefont
  {L.}~\bibnamefont {Bretheau}},\ }\bibfield  {title} {\bibinfo {title}
  {Two-tone spectroscopy of high-frequency quantum circuits with a
  {{Josephson}} emitter},\ }\href
  {https://doi.org/10.1103/PhysRevApplied.22.064027} {\bibfield  {journal}
  {\bibinfo  {journal} {Physical Review Applied}\ }\textbf {\bibinfo {volume}
  {22}},\ \bibinfo {pages} {064027} (\bibinfo {year} {2024})}\BibitemShut
  {NoStop}%
\bibitem [{\citenamefont {Albert}\ \emph {et~al.}(2024)\citenamefont {Albert},
  \citenamefont {Griesmar}, \citenamefont {Blanchet}, \citenamefont {Martel},
  \citenamefont {Bourlet},\ and\ \citenamefont
  {Hofheinz}}]{albertMicrowavePhotonNumberAmplification2024}%
  \BibitemOpen
  \bibfield  {author} {\bibinfo {author} {\bibfnamefont {R.}~\bibnamefont
  {Albert}}, \bibinfo {author} {\bibfnamefont {J.}~\bibnamefont {Griesmar}},
  \bibinfo {author} {\bibfnamefont {F.}~\bibnamefont {Blanchet}}, \bibinfo
  {author} {\bibfnamefont {U.}~\bibnamefont {Martel}}, \bibinfo {author}
  {\bibfnamefont {N.}~\bibnamefont {Bourlet}},\ and\ \bibinfo {author}
  {\bibfnamefont {M.}~\bibnamefont {Hofheinz}},\ }\bibfield  {title} {\bibinfo
  {title} {Microwave {{Photon-Number Amplification}}},\ }\href
  {https://doi.org/10.1103/PhysRevX.14.011011} {\bibfield  {journal} {\bibinfo
  {journal} {Physical Review X}\ }\textbf {\bibinfo {volume} {14}},\ \bibinfo
  {pages} {011011} (\bibinfo {year} {2024})}\BibitemShut {NoStop}%
\bibitem [{\citenamefont {Martel}\ \emph {et~al.}(2025)\citenamefont {Martel},
  \citenamefont {Albert}, \citenamefont {Blanchet}, \citenamefont {Griesmar},
  \citenamefont {Ouellet}, \citenamefont {Therrien}, \citenamefont {Nehra},
  \citenamefont {Bourlet}, \citenamefont {Peugeot},\ and\ \citenamefont
  {Hofheinz}}]{martel}%
  \BibitemOpen
  \bibfield  {author} {\bibinfo {author} {\bibfnamefont {U.}~\bibnamefont
  {Martel}}, \bibinfo {author} {\bibfnamefont {R.}~\bibnamefont {Albert}},
  \bibinfo {author} {\bibfnamefont {F.}~\bibnamefont {Blanchet}}, \bibinfo
  {author} {\bibfnamefont {J.}~\bibnamefont {Griesmar}}, \bibinfo {author}
  {\bibfnamefont {G.}~\bibnamefont {Ouellet}}, \bibinfo {author} {\bibfnamefont
  {H.}~\bibnamefont {Therrien}}, \bibinfo {author} {\bibfnamefont
  {N.}~\bibnamefont {Nehra}}, \bibinfo {author} {\bibfnamefont
  {N.}~\bibnamefont {Bourlet}}, \bibinfo {author} {\bibfnamefont
  {A.}~\bibnamefont {Peugeot}},\ and\ \bibinfo {author} {\bibfnamefont
  {M.}~\bibnamefont {Hofheinz}},\ }\bibfield  {title} {\bibinfo {title}
  {Influence of bias-voltage noise on the inelastic cooper-pair tunneling
  amplifier (icta)},\ }\href {https://doi.org/10.1063/5.0240842} {\bibfield
  {journal} {\bibinfo  {journal} {Applied Physics Letters}\ }\textbf {\bibinfo
  {volume} {126}},\ \bibinfo {pages} {074001} (\bibinfo {year}
  {2025})}\BibitemShut {NoStop}%
\bibitem [{\citenamefont {Danner}\ \emph {et~al.}(2025)\citenamefont {Danner},
  \citenamefont {H\"ohe}, \citenamefont {Padurariu}, \citenamefont
  {Ankerhold},\ and\ \citenamefont {Kubala}}]{danner}%
  \BibitemOpen
  \bibfield  {author} {\bibinfo {author} {\bibfnamefont {L.}~\bibnamefont
  {Danner}}, \bibinfo {author} {\bibfnamefont {F.}~\bibnamefont {H\"ohe}},
  \bibinfo {author} {\bibfnamefont {C.}~\bibnamefont {Padurariu}}, \bibinfo
  {author} {\bibfnamefont {J.}~\bibnamefont {Ankerhold}},\ and\ \bibinfo
  {author} {\bibfnamefont {B.}~\bibnamefont {Kubala}},\ }\bibfield  {title}
  {\bibinfo {title} {Quantum microwaves: Stabilizing squeezed light by phase
  locking},\ }\href {https://doi.org/10.1103/PhysRevB.111.184519} {\bibfield
  {journal} {\bibinfo  {journal} {Phys. Rev. B}\ }\textbf {\bibinfo {volume}
  {111}},\ \bibinfo {pages} {184519} (\bibinfo {year} {2025})}\BibitemShut
  {NoStop}%
\bibitem [{\citenamefont {Smirr}\ \emph {et~al.}(2024)\citenamefont {Smirr},
  \citenamefont {Manset},\ and\ \citenamefont {Girit}}]{smirr_2024_14577236}%
  \BibitemOpen
  \bibfield  {author} {\bibinfo {author} {\bibfnamefont {J.-L.}\ \bibnamefont
  {Smirr}}, \bibinfo {author} {\bibfnamefont {P.}~\bibnamefont {Manset}},\ and\
  \bibinfo {author} {\bibfnamefont {{\c{C}}.~{\"{O}}.}\ \bibnamefont {Girit}},\
  }\bibfield  {title} {\bibinfo {title} {Figure data for paper arxiv.2412.10227
  : Tunable josephson voltage source for quantum circuits},\ }\href
  {https://doi.org/10.5281/zenodo.14577236} {10.5281/zenodo.14577236} (\bibinfo
  {year} {2024})\BibitemShut {NoStop}%
\bibitem [{\citenamefont {Beev}\ and\ \citenamefont
  {Kiviranta}(2012)}]{beev_note_2012}%
  \BibitemOpen
  \bibfield  {author} {\bibinfo {author} {\bibfnamefont {N.}~\bibnamefont
  {Beev}}\ and\ \bibinfo {author} {\bibfnamefont {M.}~\bibnamefont
  {Kiviranta}},\ }\bibfield  {title} {\bibinfo {title} {Note: {Cryogenic}
  low-noise dc-coupled wideband differential amplifier based on {SiGe}
  heterojunction bipolar transistors},\ }\href
  {https://doi.org/10.1063/1.4729665} {\bibfield  {journal} {\bibinfo
  {journal} {Review of Scientific Instruments}\ }\textbf {\bibinfo {volume}
  {83}},\ \bibinfo {pages} {066107} (\bibinfo {year} {2012})}\BibitemShut
  {NoStop}%
\end{thebibliography}%

\end{document}